\documentclass[useAMS,usenatbib]{mn2e}
\usepackage{amsmath}
\usepackage{natbib}
\usepackage{hyperref}
\usepackage{graphicx}
\usepackage{color}
\usepackage{rotating}
\usepackage{nicefrac}
\usepackage{txfonts}
\usepackage{subfigure}
\usepackage{longtable}
\usepackage[percent]{overpic}
\usepackage{xspace}
\usepackage{url}
\usepackage{amssymb}
\usepackage[percent]{overpic}

\usepackage{epsfig}
\usepackage{amsmath}
\usepackage{rotating}
\usepackage{natbib}
\usepackage{enumerate}
\usepackage{multirow}
\renewcommand\Re{\operatorname{\mathfrak{Re}}}

\usepackage{array}
\usepackage{appendix}
\usepackage{comment}

\def \aap{A\&A}

\def \apjl{ApJ}
\def \apjs{ApJS}

\def\c#1{{#1}\xspace}

\newcommand{\ias}{{\sc \tt IAS15}\xspace}
\newcommand{\gr}{Gau\ss-Radau\xspace}
\newcommand{\radau}{{\sc \tt RADAU}\xspace}
\newcommand{\reb}{{\sc \tt REBOUND}\xspace}
\newcommand{\mer}{{\sc \tt MERCURY}\xspace}

\newcommand{\brp}[1]{\left( #1 \right)}

\def\gsim{\;\rlap{\lower 2.5pt
 \hbox{$\sim$}}\raise 1.5pt\hbox{$>$}\;}
\def\lsim{\;\rlap{\lower 2.5pt
   \hbox{$\sim$}}\raise 1.5pt\hbox{$<$}\;}

\title[\ias: A fast, adaptive, high-order integrator for gravitational dynamics]{\ias: A fast, adaptive, high-order integrator for gravitational dynamics, accurate to machine precision over a billion orbits}

\date{Accepted 2014 October 15.  Received 2014 October 14; in original form 2014 September 7}

\author[Hanno Rein \& David Spiegel]{
Hanno Rein$^{1,2}$,
David S. Spiegel$^{2}$
\\
$^1$University of Toronto, Department of Environmental and Physical Sciences, Scarborough\\
$^2$School of Natural Sciences, Institute for Advanced Study,  Princeton
}

\vspace{0.5\baselineskip}

\begin{document}
\maketitle

\begin{abstract}
We present \ias, a 15th-order integrator to simulate gravitational dynamics.
The integrator is based on a \gr quadrature and can handle conservative as well as non-conservative forces.
We develop a step-size control that can automatically choose an optimal timestep.
The algorithm can handle close encounters and high-eccentricity orbits.
The systematic errors are kept well below machine precision and long-term orbit integrations over $10^9$ orbits show that \ias is \emph{optimal} in the sense that it follows Brouwer's law, i.e. the energy error behaves like a random walk.
Our tests show that \ias is superior to a mixed-variable symplectic integrator (MVS) and other \c{popular integrators, including high-order ones,} in both speed and accuracy.
\c{In fact, \ias preserves the symplecticity of Hamiltonian systems better than the commonly-used nominally symplectic integrators to which we compared it.}

We provide an open-source implementation of \ias.
The package comes with several easy-to-extend examples involving resonant planetary systems, Kozai-Lidov cycles, close encounters, radiation pressure, quadrupole moment, and generic damping functions that can, among other things, be used to simulate planet-disc interactions.
Other non-conservative forces can be added easily.

\end{abstract}


\section{Introduction}
\label{sec:intro}
Celestial mechanics, the field that deals with the motion of celestial objects, has been an active field of research since the days of Newton and Kepler.
Ever since the availability of computers, a main focus has been on the accurate calculation of orbital dynamics on such machines.

Traditionally, long-term integrations of Hamiltonian systems such as orbital dynamics are preferentially performed using symplectic\footnote{The word `symplectic' was first proposed by \cite{Weyl1939}.} integrators \citep{Vogelaere1956,Ruth1983, Feng1985}.
\c{It is worthwhile to note that some researchers have also used symmetric, multistep methods that have good energy conservation properties (but are not symplectic) to approach this type of problem \citep{quinlan+tremaine1990}.}
Symplectic integrators have several advantages over non-symplectic integrators: they \c{conserve all the Poincar\'{e} invariants such as the the phase-space density} and have a conserved quantity that is considered a slightly perturbed version of the original Hamiltonian. 
In many situations this translates to an upper bound on the total energy error.
In the context of celestial mechanics, and if the objects of interest are on nearly Keplerian orbits, then a mixed-variable symplectic integrator is particularly useful, allowing for high precision at relatively large timesteps.
Often, a 2nd order mixed-variable symplectic integrator is sufficient for standard calculations \citep{WisdomHolman1991}.

However, there are several complications that arise when using a symplectic integrator. 
One of them is the difficulty in making the timestep adaptive while keeping the symplecticity \citep{Gladman1991,Hairer2006}.
Most mixed-variable symplectic integrators work in a heliocentric frame, and require that one object be identified as the ``star.''
This can lead to problems if there is no well-defined central object, such as in stellar binaries.
Another problem is non-conservative forces, forces that cannot be described by a potential but are important for various objects in astrophysics.
Radiation forces are a typical example of such forces, as they depend on the particle's velocity, not only its position.
Dust in protoplanetary and debris discs as well as particles in planetary rings are subject to this force.
When non-conservative forces are included in the equations of motion, the idea of a symplectic integrator --- which depends on the system being Hamiltonian --- breaks down.
\c{Within the framework of symplectic integrators, several authors, including \cite{Malhotra1994} and \cite{Mikkola1997}, provide possible solutions for this problem.}

In this paper, we take a completely different approach to orbit integrations, one that does not depend on the integrator being symplectic.
We present an integrator, the \textbf{I}mplicit integrator with \textbf{A}daptive time\textbf{S}tepping, \textbf{15}th order (\ias), that can perform calculations with high precision even if velocity-dependent forces are present. 
Due to the high order, the scheme conserves energy well below machine precision (double floating point). 

We will show that the conservation properties of \ias are as good as or even better than those of traditional symplectic integrators.
The reason for this is that, in addition to errors resulting from the numerical scheme, there are errors associated with the finite precision of floating-point numbers on a computer.
All integrators suffer from these errors.
\c{\citet{Newcomb1899} was one of the first to systematically study the propagation of such errors \citep[see also][]{Schlesinger1917, Brouwer1937, Henrici1962}.}
If every operation in an algorithm is unbiased (the actual result is rounded to the nearest representable floating-point number), then the error grows like a random walk, i.e.~$\propto n^{1/2}$, where~$n$~is the number of steps performed.
Angle-like quantities, such as the phase of an orbit, will grow faster, ~$\propto n^{3/2}$. 
This is known as Brouwer's law.

We will show that \ias is \emph{optimal} in the sense that its energy error follows Brouwer's law. 
Unless extended precision is used, an integrator cannot be more accurate.\footnote{The accuracy is limited by the precision of the particles' position and velocity coordinates at the beginning and end of each timestep.}
\c{Popular implementations of other integrators} used for long-term orbit integrations, symplectic or not, do not follow Brouwer's law and show a linear energy growth in long simulations.
This includes, but is not limited to the standard \citeauthor{WisdomHolman1991} (WH) integrator, Mixed-Variable Symplectic (MVS) integrators with higher order correctors, Bulirsch-Stoer integrators~(BS) and \gr integrators.\footnote{We test the implementations provided by the MERCURY package \citep{Chambers1997, chambers1999}. Consistent results have been obtained using other implementations \citep[e.g.][]{ReinLiu2012}.}
It is worth pointing out yet again that this happens despite the fact that WH and MVS are symplectic integrators, which are generally thought of as having a bounded scheme error, but no such guarantee about errors associated with limited floating-point precision can be made.

Because \ias is a very high order scheme, we are able to perform long-term simulations with machine precision maintained over at least~$10^9$~orbital timescales using only 100~timesteps per orbit. 
This is an order of magnitude fewer timesteps than required by NBI, another integrator that achieves Brouwer's law \citep{Grazier2005}.

Whereas the \gr scheme itself had been used for orbital mechanics before \citep{Everhart1985,Chambers1997, Hairer2008}, we develop three approaches to significantly improve its accuracy and usefulness for astrophysical applications.
First, we show that the integrator's systematic error can be suppressed to well below machine precision.
This makes it possible to use \ias for long-term orbit integrations.
Second, we develop a new automatic step-size control based on a novel physical interpretation of the error terms in the scheme.
Contrary to previous attempts, our step-size control does not contain any arbitrary (and unjustified) scales.
Third, we ensure, using various techniques such as compensated summation, that the round-off errors are symmetric and at machine precision.

We implement \ias into the freely available particle code \reb, available at \url{http://github.com/hannorein/rebound}. 
Due to its modular nature, \reb can handle a wide variety of astrophysical problems such as planetary rings, granular flows, planet migration and long-term orbit integration \citep{ReinLiu2012}.
\c{We also provide a simple python wrapper.}

The remainder of this paper is structured as follows.
We first present the principle of a \gr integrator and our implementation thereof, the \ias integrator in Section~\ref{sec:RADAU}.
We point out several significant improvements that we added, including the adaptive timestepping.
In Section~\ref{sec:errors} we estimate the error associated with the integrator and compare it to the floating-point error.
This is also where we point out how to keep round-off errors symmetric and at machine precision.
Together, these results can then be used to ensure that \ias follows Brouwer's law.
We test our integrator in a wide variety of cases in Section~\ref{sec:tests} before summarizing our results in Section~\ref{sec:conclusions}.
In appendix \ref{sec:PR}, we present a simple (and, in our view intuitive) derivation of the Poynting-Robertson drag \citep{Poynting1903, Robertson1937} which is used in one of the example problems \c{that we include in the public distribution of \ias.}

%%%%%%%%%%%%%%%%%%%%%%%%%%%%%%%%%%%%%%%%%%%%%%%%%%%%%
\section{\ias Integrator}
\label{sec:RADAU}

\citet{Everhart1985} describes an elegant 15th-order modified Runge-Kutta integrator.
\c{His work is an important contribution to celestial mechanics that makes highly accurate calculations possible.} 
\c{Our work builds on his success.}
We implemented Everhart's algorithm in C99 and added it to the \reb code \c{(a python wrapper is also provided)}.
In addition, we fixed several flaws and improved the algorithm so that its accuracy stays at machine precision for billions of dynamical times. 
The ORSA toolchain \citep{ORSA} was partly used in this development. 
We improved both the accuracy and the step size control in our implementation, which we refer to as \ias (Implicit integrator with Adaptive timeStepping, 15th~order). 
Here, we summarize the underlying \gr algorithm, so as to provide the reader with sufficient context, and we then point out our modifications.

To avoid confusion, we use square brackets $[\;]$ for function evaluations and round brackets $(\;)$ for ordering operations in the remainder of the text.

\subsection{Algorithm}
The fundamental equation we are trying to solve is 
\begin{eqnarray}
y'' &=& F[y',y,t], \label{eq:fund}
\end{eqnarray}
where $y''$ is the acceleration of a particle and $F$ is a function describing the specific force, which depends on the particle's position $y$, the particle's velocity $y'$, and time $t$.
This equation is general enough to allow arbitrary velocity-dependent and therefore non-conservative forces.
Thus, the system might not correspond to a Hamiltonian system.

We first expand equation (\ref{eq:fund}) into a truncated series:
\begin{eqnarray}
\label{eq:a_equation} y''[t] & \approx & y_0'' + a_0 t + a_1 t^2 + \ldots + a_6 t^7.
\end{eqnarray}
The constant term in the above equations is simply the force at the beginning of a timestep, $y_0''\equiv  y''[0] = F[t=0]$.
Introducing the step size $dt$ as well as $h\equiv t/dt$, and $b_k \equiv a_k dt^{k+1}$, we can rewrite the expansion as
\begin{eqnarray}
\label{eq:b_equation} y''[h] & \approx & y_0'' + b_0 h + b_1 h^2 + \ldots + b_6 h^7.
\end{eqnarray}
Note that, since $h$ is dimensionless, each coefficient $b_k$ has dimensions of acceleration.
Before moving on, let us rewrite the series one more time in the form:
\begin{eqnarray}
\label{eq:g_equation} y''[h] & \approx & y''_0 + g_1h + g_2h(h-h_1) + g_3 h(h-h_1)(h-h_2) \\
&&+ \ldots + g_8 h (h-h_1) \ldots (h-h_7).\nonumber
\end{eqnarray}
The newly introduced coefficients~$g_k$ can be expressed in terms of~$b_k$ and vice versa by comparing equations (\ref{eq:b_equation}) and (\ref{eq:g_equation}).
For now,~$h_1, \ldots, h_7$ are just coefficients in the interval $[0,1]$.
We later refer to them as substeps (within one timestep).
The advantage of writing the expansion this way is that a coefficient $g_k$  depends only on the force evaluations at substeps $h_n$ with $n\leq k$, for example:
\begin{eqnarray}
h=h_1 \quad\text{ gives } \quad && g_1=\frac{y''_1-y''_0}{h_1}\nonumber\\ 
h=h_2 \quad\text{ gives } \quad && g_2=\frac{y''_2-y''_0 -g_1h_2}{h_2(h_2-h_1)}\label{eq:g}\\
\ldots\nonumber
\end{eqnarray}
where we introduced $y_n'' \equiv y''[h_n]$.
In other words, by expressing $y''$ this way, when we later step through the substeps $h_n$, we can update the coefficients $g_k$ on the fly.
Each $b_k$ coefficient, on the other hand, depends on the force evaluations at all substeps.

So far, we have only talked about the approximation and expansion of the second derivative of $y$, the acceleration.
What we are actually interested in are the position and velocity, $y$ and $y'$.
Let us therefore first integrate equation (\ref{eq:b_equation}) once to get an estimate of the velocity at an arbitrary time during and at the end of the timestep:
\begin{eqnarray}
\label{eq:v_equation} y'[h] & \approx & y'_0 +h\,dt \,\left(y_0'' + \frac{h}{2} \left( b_0+ \frac{2h}{3}\Bigg( b_1+ \ldots\Bigg)\right)\right)
\end{eqnarray}
If we integrate the result once again, we get an estimate of the positions at an arbitrary time during and at the end of the timestep:
\begin{eqnarray}
\label{eq:y_equation} y[h] & \approx & y_0 + y'_0\,h\,dt + \frac{h^2 dt^2}{2} \,\left(y_0''+\frac{h}{3}\,\left(b_0+\frac{h}{2}\,\Bigg(b_1 + \ldots\Bigg)\right)\right).
\end{eqnarray}
The first two terms correspond to the position and velocity at the beginning of the timestep.
The trick to make the approximation of this integral very accurate is to choose the spacing of the substeps to be \gr spacing (rather than, for example, equidistant spacing).
\gr spacing is closely related to the standard Gau{\ss}ian quadrature which can be used to approximate an integral, but makes use of the starting point at $h=0$ as well (whereas standard Gaui{\ss}ian quadrature only uses evaluation points in the interior of the interval).
This gives us an advantage because we already know the positions and velocities of the particles at the beginning of the timestep.
We use a quadrature with~$8$~function evaluations to construct a 15th-order scheme.\footnote{The numerical values of the spacings including only the four most significant digits are $h_n=0$, $0.0562$, $0.1802$, $0.3526$, $0.5471$, $0.7342$, $0.8853$ and $0.9775$. The constants are roughly the same as given by \cite{Everhart1985} but we implement them with more than 16 decimal places to avoid any bias in converting them to a floating point number.}

We now \textit{only} have to find good estimates for the coefficients~$b_k$ to get the positions and velocities via equations (\ref{eq:v_equation}) and (\ref{eq:y_equation}).
To do this, we need estimates of the forces during the timestep, which we take at the substeps $h_n$, the same sub-step times that we later use to approximate the integrals.
The force estimates give us the $g_k$ coefficients via equation (\ref{eq:g}), which we can then convert to the $b_k$ coefficients.
This is of course an implicit system.
The forces depend on the a priori unknown positions and velocities.

We solve this dilemma with a predictor corrector scheme:
First, we come up with a rough estimate for the positions and velocities to calculate the forces (predictor).
In the very first iteration, we simply set all $b_k=0$, corresponding to a particle moving along a path of constant acceleration.
Then, we use the forces to calculate better estimates for the positions and velocities (corrector).
This process is iterated until the positions and velocities converged to machine precision.
Below, we describe in detail how we measure the convergence.

The only time we actually have to set the $b_k$ values to zero is the very first timestep.
In any subsequent timestep, we can make use of previous knowledge, as the $b_k$ coefficients are only slowly varying.
Better yet, we can compare the final values of $b_k$ to the predicted value.
This correction, $e_k$, can then be applied to the $b_k$ predictions for the next timestep, making the predicted values even better.
Only very few ($\sim 2$) iterations of the predictor corrector loop are needed to achieve machine precision.

%----------------------------------------------------
\subsection{Warmup procedure and convergence of the predictor-corrector loop}
In the original implementation by Everhart, it was suggested to use a fixed number of six iterations for the predictor corrector loop in the first timestep and two thereafter.
This statement was based on experimentation and experience.
Here we present a better, quantitative approach to determining the number of iterations.

However, before we present our approach, it is worth pointing out that the procedure by \cite{Everhart1985} can be improved in another, very simple way, by using six iterations for both the first \emph{and} the second timestep.
While we only need one timestep to warmup and be able to use knowledge from the previous timestep to get better $b_k$ values, we need two timesteps to capture the slowly varying corrections $e_k$, thus the increased number of iterations in the first two timesteps.

While the above procedure does result in an improvement, we can do even better.
In our implementation we do not set a predefined number of predictor-corrector iterations, but determine it dynamically.
At the end of each iteration, we measure the change made to the coefficient $b_6$ relative to the acceleration $y''$.
We implement two different ways to calculate this ratio.
\begin{equation}
\widetilde{\delta b_6} = \left\{
	\begin{array}{ll}
	\frac{\displaystyle {\max}_i\,{\left|\delta b_{6,i}\right|}}{\displaystyle{\max}_i\,{\left|y''_i\right|}} & 
		\mbox{ global error estimate}\\\\
	\max_i\left|\frac{\displaystyle{\delta b_{6,i}}}{\displaystyle y''_i}\right|&
		\mbox{ local error estimate}
	\end{array} 
	\right.
 \label{eq:globallocal}
\end{equation}
where the index $i$ runs over all three \c{components of all particles}. 
By default, we use the ratio that we call the global error estimate.
This choice is preferred in the vast majority of situations and gives a reasonable estimate of the error.\footnote{As a counter example, consider the following hierarchical system.
A moon is on an eccentric orbit around a planet.
Another planet is on an extremely tight orbit around the star.
Here, the error in $b_6$ could be dominated by the moon, whereas $y''$ could be dominated by the planet on the tight orbit.
In that case $\epsilon_{\rm local}$ will give an improved (smaller) error estimate.
The user can switch to the local measure by setting the variable \texttt{integrator\_epsilon\_global} to zero.}
When the series has converged to machine precision, the change to $b_6$ is insignificant compared to $y''$, i.e. $\widetilde{\delta b_6} < \epsilon_{\delta b}$, where we choose $\epsilon_{\delta b} \equiv 10^{-16}$).
We further terminate the predictor-corrector loop if $\widetilde{\delta b_6}$ begins to oscillate as it is an indication that future iterations are unlikely to improve the accuracy.
Note that the value $\epsilon_{\delta b}$ does not determine the final order of the scheme or even the accuracy.
It merely ensures that the implicit part of the integrator has converged.

In most cases the iteration converges with only 2 iterations.\footnote{Everhart's guess was quite reasonable.}
However, there are cases where more steps are required.
For example, during the initial two timesteps and during any sudden changes in the system~(such as a close encounter).

To prevent infinite loops, we set an upper limit of 12 iterations.
If the iteration has still not converged, the timestep is almost certainly too large.
As an example, imagine a situation where the timestep is 100 times larger than the orbital period.
No matter how many iterations we perform, we cannot capture the relevant timescale.
In such a case, a warning message alerts the user.
As long as adaptive timestepping is turned on (see below), one need not worry about such a scenario, because the timestep is automatically chosen so that it is smaller than any physical timescale.

%----------------------------------------------------
\subsection{Stepsize control}
\label{sec:stepsize}
We now move on to present a new way to automatically choose the timestep.
This is different from and superior to the one proposed by \cite{Everhart1985}.
The precision of the scheme is controlled by one dimensionless parameter $\epsilon_b\ll1$. 
In a nutshell, this parameter controls the stepsize by requiring that the function~$y[t]$ be smooth within one timestep.
We discuss in Section~\ref{sec:errors} what this means in detail and how this value is chosen to ensure that the timestep is smaller than the typical timescale in the problem.
Note that \cite{Everhart1985} uses a dimensional parameter for his \radau integrator, dramatically limiting its usefulness and posing a potential pitfall for its use.
A simple change of code units can make \radau fail spectacularly. 

For a reasonable\footnote{The timestep has to be comparable to the shortest relevant timescale in the problem.} step size $dt$, the error in the expansion of $y''$ will be smaller than the last term of the series evaluated at $t=dt$ or $h=1$; i.e., the error will not be larger than $b_6$.  
An upper bound to the relative error in the acceleration, is then $\widetilde{b_6} = b_6/y''$.  
We calculate this ratio globally by default, but offer an option to the user to use a local version instead:
\begin{equation}
\widetilde{b_6} = \left\{
	\begin{array}{ll}
	\frac{\displaystyle {\max}_i\,{\left|b_{6,i}\right|}}{\displaystyle{\max}_i\,{\left|y''_i\right|}} & 
		\mbox{ global error estimate}\\\\
	{\max_i}\,\left|\frac{\displaystyle{b_{6,i}}}{\displaystyle y''_i}\right|&
		\mbox{ local error estimate}
	\end{array} 
	\right.
\label{eq:globallocal2}
\end{equation}
Note that these expressions are similar to those in equation~(\ref{eq:globallocal}).
But here, instead of using the change to the $b_{6,i}$ coefficients, we use their values.  

By comparing equations~(\ref{eq:a_equation}) and~(\ref{eq:b_equation}) one finds that $a_6t^7 = b_6h^7 = b_6t^7/dt^7$.
Thus changing the timestep by a factor $f$ will change $b_6$ by a factor of $f^{7}$.  
In other words, for two different trial timesteps  $dt_A$ and $dt_B$ the corresponding fractions $\widetilde{b_6}_A$ and $\widetilde{b_6}_B$ are related as
\begin{eqnarray}
\label{eq:error_ratio} \widetilde{b_6}_A dt_B^{7} = \widetilde{b_6}_B dt_A^{7} \, .
\end{eqnarray}
To accept an integration step, we perform the following procedure. 
First, we assume a trial timestep $dt_{\rm trial}$.
We use the initial timestep provided by the user in the first timestep.
In subsequent timesteps, we use the timestep set in the previous step.
After integrating through the timestep (and converging the predictor corrector iteration, see above), we calculate~$\widetilde{b_6}$.  
The accuracy parameter~$\epsilon_b$ is then used to calculate the required timestep as
\begin{eqnarray}
\label{eq:error_dt} dt_{\rm required} = dt_{\rm trial} \cdot \left( {\epsilon_b}\,/\,{\widetilde{b_6}} \;\right)^{1/7}  \, .
\end{eqnarray}
If $dt_{\rm trial} > dt_{\rm required}$  the step is rejected and repeated with a smaller timestep. 
If the step is accepted, then the timestep for the next step is chosen to be $dt_{\rm required}$. 

Because we use a 15th-order integrator to integrate $y''$, the actual relative energy error (involving $y$ and $y'$) can be many orders of magnitude smaller than the accuracy parameter $\epsilon_b$.
We come back to how to choose $\epsilon_b$ in the next section.

Our implementation has the advantage that the stepsize chosen is independent of the physical scale of the problem, which is not the case in the original implementation of \cite{Everhart1985}. 
For instance for an equal mass binary with total mass $M$ and separated by a distance $a$, rescaling the problem in a way that leaves the dynamical time constant (constant $a^3/M$) also leaves the timestep~$dt$ unchanged.
In contrast, in Everhart's implementation, the error is constrained to be less than a dimensional length, so rescaling a problem to smaller mass and length scales at fixed dynamical time results in longer timesteps and positional errors that are a larger fraction of the scale of the problem.

It is worth re-emphasizing that our scheme is 15th~order, so shrinking the timestep by a factor of $\alpha$ can shrink the error by a factor of $\alpha^{16}$.
Therefore, there is only one decade in $dt$ between being correct up to machine precision and not capturing the physical timescale at all~(see below).
The procedure above ensures that we are capturing the physical timescales correctly.
One can of course always construct a scenario where this method will also fail.
The sort of scenario would involve a change in the characteristic timescale of the system over many e-foldings happening within a single timestep.
We have not been able to come up with any physical scenario where this would be the case.

Note the difference between the statement in the previous paragraph and equation~(\ref{eq:error_ratio}), where the exponent is 7, not 16.  
The exponents are different because there are two different quantities that we wish to estimate.  
One task is to estimate the $b$ coefficients as accurately as possible.  
The other task is to estimate the error in the integration, given the $b$ coefficients.  
Even if the $b$ coefficients were known exactly, the integral would be an approximation, and, owing to its 15-order nature, the error from integrating the force would scale with $dt^{16}$. 

%%%%%%%%%%%%%%%%%%%%%%%%%%%%%%%%%%%%%%%%%%%%%%%%%%%%%
\section{Error estimates}
\label{sec:errors}
It is notoriously difficult to define how \emph{good} an integrator is for any possible scenario because the word ``good'' is ambiguous and can mean different things in different contexts.
We focus the following discussion on the energy error, but briefly mention the phase error.
Other error estimates such as velocity and position error might have different scalings. 
We also focus on the two-body problem, a simple test case where we know the correct answer.
Realistic test cases are shown in Section~\ref{sec:tests}.

We first discuss what kind of errors are expected to occur, their magnitude, and their growth.
Then, in Section \ref{sec:errias}, we derive explicitly an error estimate for \ias.
This error estimate is used to set the precision parameter $\epsilon_b$.

We only consider schemes with constant timestep in this Section.
The error growth is described in terms of physical time $t$, which is equivalent to $N\cdot dt$, where $N$ is the number of integration steps with timestep $dt$.

%----------------------------------------------------
\subsection{Machine precision}
We are working in double floating-point precision IEEE 754 (see specification ISO/IEC/IEEE 60559:2011).
A number stored in this format has a significand precision of 52 bits, corresponding to about 16 decimal digits. 
Thus, any relative error that we compute can only be accurate to within $\sim$2$\cdot 10^{-16}$.  
We call this contribution to the total error $E_{\rm floor}$.

%----------------------------------------------------
\subsection{Random errors}
Any\footnote{There are a few exceptions such as 2.+2. or 0. times anything.} calculation involving two floating-point numbers will only be approximate when performed on a computer.
The IEEE~754 standard ensures that the error is randomly distributed. 
For example, the addition of two random floating-point numbers $a$ and $b$ will add up to a number $a+b+\epsilon$ with the error $\epsilon$ being positive or negative with equal probability. 
As the simulation progresses, this contribution to the error will grow.
If the contribution for each individual calculation is random, the quantity will grow as $\propto t^{1/2}$. 
Some quantities are effectively an integral of another quantity over time, such as the mean longitude. 
Such quantities (angles) accumulate error faster and grow as $\propto t^{3/2}$.
These relations are known as Brouwer's law \citep{Brouwer1937}.
Let us call this contribution to the error $E_{\rm rand}$.
\ias makes use of different concepts which we explain in the following to ensure a small random error $E_{\rm rand}$. 

First, a concept called compensated summation \citep{Kahan1965,Higham2002,Hairer2006} is used.
The idea is to improve the accuracy of additions that involve one small and one large floating-point number.
Such a scenario naturally occurs in all integrators for example when updating the positions:
\begin{eqnarray}
x_{n+1} = x_{n} + \delta x.
\end{eqnarray}
In the above equation, the change $\delta x$ is typically small compared to the quantity $x$ itself.
Thus, many significant digits are lost if implemented in the most straight-forward way.
Compensated summation enables one to keep track of the lost precision and therefore reduce the buildup of random errors, effectively working in extended precision with minimal additional work for the CPU.
We implemented compensated summation for all updates of the position and velocity at the end and during (sub-)timesteps.
Our experiments have shown that this decreases $E_{\rm rand}$ by one to two orders of magnitude at almost no additional cost.

Second, we made sure every constant used in \ias is as accurate as possible. 
All hard coded constants have been pre-calculated in extended floating-point precision using the GNU MP library \citep{Fousse2007}. 
In addition, multiplications of a number $x$ with a rational number $p/q$ are implemented as $p \cdot x / q$.

Third, the use of adaptive timesteps turns out to be beneficial to ensure a purely random rounding error distribution (see next section). 

It is important to point out that there are computer architectures/compilers that do not follow the rounding recommendations of the IEEE~754 standard, for example some GPU models.
If using the GNU gcc compiler, the user should \emph{not} turn on the following compiler options: \texttt{-fassociative-math}, \texttt{-freciprocal-math} and \texttt{-ffast-math}.
This can lead to an increased random error or, in the worst case, a biased random walk (see next section).

\subsection{Biased random errors}
Depending on hardware, compiler, and specific details on the implementation, the error resulting from floating-point operations might be biased and grow linearly with time.
A simple example is the repeated rotation of a vector around a fixed axis.
If the rotation is implemented using the standard rotation matrix and the rotation is performed with the same angle every time, the length of the vector will either grow or shrink linearly with time as a result of floating-point precision.
In a nutshell, the problem is that $\sin^2 \phi+\cos^2 \phi$ is not exactly~1.
Solutions to this specific problem are given by \cite{QuinnTremaine1990}, \cite{HenonPetit1998} and \cite{ReinTremaine2011}.
The latter decompose the rotation into two shear operations, which guarantees the conservation of the vector's length even with finite floating-point precision.
Let us call this contribution to the error~$E_{\rm bias}$.

It may not be possible to get rid of all biases and prove that the calculation is completely unbiased in complex algorithms.
However, we note that \ias does not use any operation other than $+, -, \cdot$~and~$/$.
In particular, we do not, \c{other than in the force evaluations}, use any square root, sine or cosine functions. 
Other integrators, for example mixed-variable symplectic integrators, do require many such function evaluations per timestep.\footnote{An extra benefit of not using these complex functions is an increase in speed.}

One interesting effect we noticed is that roundoff errors are more random (less biased) for simulations in which adaptive timestepping is turned on. 
Several constants are multiplied with the timestep before being added to another number, see e.g. equation~(\ref{eq:y_equation}). 
If the step size is exactly the same every timestep, the rounding error will be more biased than if the timestep fluctuates a small amount. 
This is counter to the effect of adaptive timestepping in symplectic integrators, where it in general increases the energy error unless special care is taken to ensure the variable timestep does not break symplecticity. 
Our integrator is not symplectic, and the scheme error is much smaller than the machine precision, so we can freely change the timestep without having a negative effect on long-term error growth.
For these reasons, we encourage the user to always turn on adaptive timestepping.

We show later in numerical tests that our implementations seems to be indeed unbiased in a typical simulation of the Solar System over at least $10^{11}$ timesteps, equivalent to one billion ($10^9$)~orbits.

%----------------------------------------------------
\subsection{Error associated with the integrator}
\begin{table*}
\caption{Time dependence of the error contribution $E_{\rm scheme}$ for various integrators in the two and three body problem. 
Note that there are other contributions to the total error which might dominate.
\c{$C$ is a constant which may dependent on the specific problem but is independent of time.}
\label{tab:growth}}
\begin{tabular}{l|l|ll|l}
& & \multicolumn{2}{c|}{Two-body problem} 				& Three-body problem \\
&Integrator 			& Energy error 		& Phase error 	& Energy Error \\\hline\hline
\multirow{5}{*}{$E_{\rm scheme}$} & RK4 / BS 				& $\propto t$		& $\propto t^2$ & $\propto t$\\
& Leap Frog 			& $\leq C$ (probably bounded)	& $\propto t$	& $\leq C$ (probably bounded) \\
& WH / MVS			& $0^{*}$ 		& $0^{*}$	& $\leq C$ (probably bounded) \\
& \ias				& $\propto t$		& $\propto t^2$	& $\propto t$ \\
& \radau				& $\propto t$		& $\propto t^2$	& $\propto t$ \\\hline
$E_{\rm rand}$& Any	& $\propto t^{1/2}$ & 	$\propto t^{3/2}$ &  $\propto t^{1/2}$\\\hline 
$E_{\rm bias}$& Implementation dependent	& $\propto t$ & 	$\propto t^{2}$ &  $\propto t$ 
\end{tabular}

\vspace{.25cm}
\raggedright
$^{*}$ This ignores floating-point precision and implementation specific errors. 
\end{table*}

The error contributions $E_{\rm floor}$, $E_{\rm rand}$ and in part $E_{\rm bias}$ are inherent to all integrators that use floating-point numbers. 

In addition to these, the integrators have an error associated with themselves. 
This error contribution\footnote{This is sometimes referred to as truncation or discretization error.}, which we call $E_{\rm scheme}$, has different properties for different integrators and is the one many people care about and focus on when developing a new integrator. 
These properties depend on the quantity that we use to measure an error. 
For example, for one integrator, the energy might be bounded but the phase might grow linearly in time.

Symplectic integrators such as leapfrog and Wisdom-Holman have been shown to maintain a bounded energy error in many situations.
The energy error in non-symplectic integrators typically grows linearly with time. 
Table~\ref{tab:growth} lists how the $E_{\rm scheme}$ error grows with time for different types of integrators.
Note that the table lists only the proportionality, not the coefficients of proportionality, and therefore cannot answer which error dominates.

%----------------------------------------------------
\subsection{Total error}
\label{ssec:totalerror}
The total error of an integrator is the sum of the four contributions discussed above:
\begin{eqnarray}
E_{\rm tot} = E_{\rm floor} + E_{\rm rand} + E_{\rm bias} + E_{\rm scheme}.
\end{eqnarray}
If we define the goodness of a scheme by the size of the error, then how good a scheme is is determined by the magnitude of the largest term.
Which one that is depends on the scheme, the problem studied and the number of integration steps.
In general, there will be at least one constant term, e.g. $E_{\rm floor}$ or $E_{\rm scheme}$, and one term that grows as $t^{1/2}$, e.g. $E_{\rm rand}$.
Even worse, some terms might grow linearly with $t$, e.g. $E_{\rm scheme}$ or $E_{\rm bias}$. 
For some quantities, there might be even higher power terms depending on the integrator (see Table~\ref{tab:growth}). 

The interesting question to ask is which error dominates. 
Traditional schemes like RK4, leapfrog and Wisdom-Holman do not reach machine precision for reasonable timesteps.
In other words, $E_{\rm scheme} \gg 10^{-16}$. 
An extremely high order integrator, such as \ias, on the other hand, reaches machine precision for timesteps just an order of magnitude below those at which relative errors are order unity (i.e. where the integrator is unusable).
$E_{\rm scheme}$ becomes negligible as long as it remains below machine precision for the duration of the integrations. 
Then $E_{\rm floor}$ and $E_{\rm rand}$ (and possibly $E_{\rm bias}$) completely dominate the error.
If that is the case, \ias will always be at least equally accurate, and in most cases significantly more accurate than RK4, leapfrog and Wisdom-Holman.
In that sense, we call \ias \emph{optimal}.
It is not possible to achieve a more accurate result with a better integrator without using extended floating-point precision.

In the following, we estimate $E_{\rm scheme}$ for \ias and show that our new adaptive timestepping scheme ensures that it is negligible to within machine precision for at least $10^9$ dynamical timescales.

%----------------------------------------------------
\subsection{Error for \ias}
\label{sec:errias}
Let us now try to estimate the error $E_{\rm scheme}$ for \ias. 
We consider the error made by integrating over a single timestep, thus giving us an estimate of the error in the position $y$. 
Although we might ultimately be interested in another error, for example the energy error, the discussion of the position error will be sufficient to give an estimate of~$E_{\rm scheme}$. 

In a \gr quadrature integration scheme, such as the one we use for \ias, a function $F[t]$ is integrated on the domain $[0,dt]$ with $m$ quadrature points (i.e., $m-1$ free abscissae).
The absolute error term is given by \citet{Hildebrand1974} as
\begin{eqnarray}
\label{eq:RadauE}  =   \frac{m \left\{ (m-1)! \right\}^4}{2 \left\{ (2m-1)!\right\}^3} F^{(2m-1)}[\xi] \left( dt \right)^{2m} \, ,
\end{eqnarray}
for some $\xi$ in the domain of integration, where $F^{(2m-1)}$ represents $(2m-1)$th derivative of $F$.
\ias uses 7 free abscissae, thus we have $m=8$.
The absolute error of $y$ (a single time-integration of~$y'$) can therefore be expressed by
\begin{eqnarray}
E_{y}  =  \mathcal{C}_8 \;y^{(16)}[\xi]\; dt^{16} \quad\quad\quad \xi\in[0,dt].
\end{eqnarray} 
Here, $y^{(16)}$ is the 16th derivative of $y$ (and therefore the 15th derivative of $F=y'$).
The constant factor $\mathcal{C}_8$ can be expressed as a rational number 
\begin{eqnarray}
\mathcal{C}_8 = 1/{866396514099916800000}\approx 1.15\cdot 10^{-21} \, .
\end{eqnarray}
So far we have only an expression for the error that depends on the sixteenth derivative of $y$.
That is not of much use, as we have no reliable estimate of this derivative in general.
To solve this issue, we could switch to a nested quadrature formula such as Gau{\ss}-Kronrod to get an estimate for the error \citep{Kronrod1965}.

However, we are specifically interested in using the integrator for celestial mechanics and therefore use the following trick to estimate the magnitude of the high order derivate.
We assume that the problem that we are integrating can be decomposed in a set of harmonics.
Even complicated problems in celestial mechanics, such as close encounters, can be expressed in a series of harmonics.
This is the foundation of many perturbation theories.
For simplicity, we assume that there is only one harmonic and therefore one characteristic timescale.
This corresponds to the circular two-body problem.
More complicated problems are a generalization thereof with higher order harmonics added.

Now, consider an harmonic oscillator (this corresponds to a circular orbit) and assume the position evolves in time as
\begin{eqnarray}
y[t] = \Re\left[ y_0 e^{i\omega t}\right]\label{eq:y}
\end{eqnarray}
By introducing $\omega$, we have effectively introduced  $T={2\pi}/\omega$, the characteristic timescale of the problem. 
For example, if we consider nearly circular orbits, then this corresponds to the orbital period. 
If we consider close encounters, then this corresponds to the encounter time.
For this harmonic oscillator we can easily calculate the $n$-th derivative
\begin{eqnarray}
y^{(n)}[t] =  \Re\left[ (i \omega)^{n} y[t] \right]
\end{eqnarray}
Setting $n=16$ and taking the maximum of the absolute value gives us now an estimate of the absolute integrator error
\begin{eqnarray}
E_{y} \approx \mathcal{C}_8 \;\omega^{16} \,dt^{16} y_0 \, .
\end{eqnarray} 
The relative error $\widetilde E_{y}$, is then simply
\begin{eqnarray}
\widetilde E_{y} = \frac{E_{y}}{y_0} \approx \mathcal{C}_8 \;\omega^{16} \,dt^{16} \label{eq:ey2}
\end{eqnarray}
Let us now come back to the previously defined precision parameter $\epsilon_b$, see equation~(\ref{eq:globallocal2}).
We can express this parameter in terms of derivates of $y$ to get 
\begin{eqnarray}
\epsilon_b \equiv \frac{b_6}{y''} = \frac{a_6\,dt^7}{y''} \approx \frac{y^{(9)} dt^7}{7! \, y''} \, ,
\end{eqnarray}
The function $y$ is by assumption of the form given in equation~(\ref{eq:y}) which allows us to evaluate the expression for $\epsilon_b$ explicitly:
\begin{eqnarray}
\epsilon_b\approx \frac{\omega^7 \,dt^7}{7!}\approx \left(\frac{dt}{T}\right)^7\frac{(2\pi)^7}{7!} \, .
\label{eq:epsilon_0}
\end{eqnarray}
Solving for $\omega$ in equation~(\ref{eq:epsilon_0}) and substituting in equation~(\ref{eq:ey2}) gives
\begin{eqnarray}
\widetilde E_{y} =  3.3\cdot10^{-13}\;\epsilon_b^{16/7} 
\end{eqnarray}
We now have a relation on how the relative error over one timestep depends on the precision parameter $\epsilon_b$. 
For example, if we want to reach machine precision ($10^{-16}$) in $\widetilde E_{y}$, we need to set $\epsilon_b\approx 0.028$.
Let us finally look at the length of the timestep compared to the characteristic dynamical time $T$. 
This ratio, let us call it $\widetilde{dt}$, can also be expressed in terms of $\epsilon_b$:
\begin{eqnarray}
 \widetilde{dt} \equiv \frac{dt}{T} \approx \epsilon_b^{1/7} \frac{(7!)^{1/7}}{2\pi}
 \approx \frac1{1.86} \epsilon_b^{1/7}.
\end{eqnarray}
With $\epsilon_b\approx 0.028$, we have $\widetilde{dt}\approx 0.3$.
The implications of setting $\epsilon_b$ to $0.028$ are thus two-fold.
First, we reach machine precision in $\widetilde E_{y}$. 
Second, we have ensured that the timestep is a fraction of the smallest characteristic dynamical time. 

The above estimate ensures that we reach machine precision after one timestep.
However, we want to further decrease the timestep for two reasons.
First, if we use the exact timestep estimate from above, we ignore that the timestep estimate might vary during the timestep (not by much, but noticeably so).
If the timestep criterion is not satisfied at the end of the timestep, we would need to repeat the timestep.
If this happens too often, then we end up doing more work than by just using a slightly smaller timestep to start with.
Second, although setting $\epsilon_b=0.028$ ensures that we reach machine precision after one timestep, we also want to suppress the long-term growth of any systematic error in $E_{\rm scheme}$.
If we consider a total simulation time of $10^{10}$ orbits, the error per timestep needs to be smaller by a factor of $10^{10}\cdot \widetilde{dt}{}^{-1}$ assuming a linear error growth (i.e. divided by the number of timesteps).
In a typical simulation with moderately eccentric orbits, we will later see that $\widetilde{dt}\sim0.01$, giving us a required reduction factor of $10^{12}$.
Luckily, we have a high order scheme.
To reduce the scheme error after one timestep by a factor of $10^{12}$, i.e. reaching $E_{\rm scheme} \sim 10^{-28}$~(!), we only need to reduce the timestep by a factor of 6. 
Putting everything together we arrive at a value of $\epsilon_b \approx 10^{-7}$. 
We take a slightly conservative value and set a default value of $\epsilon_b = 10^{-9}$, which corresponds to a timestep a factor of 2 smaller than our error estimation suggests.

The user can change the value of $\epsilon_b$ to force the integrator to use even smaller or larger timesteps. 
This might be useful in certain special circumstances, but in most simulations the error might actually increase with an increase in the number of timesteps as the random errors dominate and accumulate (see above).
It is also worth emphasizing again that $\epsilon_b$ is not the precision of the integrator, it is rather a small dimensionless number that ensures the integrator is accurate to machine precision after many dynamical timescales.

Now that we have derived a physical meaning for the $\epsilon_b$ parameter, let us have another look at equation~\ref{eq:globallocal2}.
This ratio might not give a reliable error estimate for some extreme cases.
The reason for this is the limited precision of floating point numbers. 
For the ratio in equation~\ref{eq:globallocal2} to give a reasonable estimate of the smoothness of the acceleration, we need to know the acceleration accurately enough. 
But the precision of the acceleration can at best be as good as the precision of the relative position between particles.
For example, if the particle in question is at the origin, then the precision of the position (and therefore the acceleration) is roughly $E_{\rm floor}$,  $10^{-16}$.
However, if the particle is offset, then the precision in the relative position is degraded.
Imagine an extreme example, where the Solar System planets are integrated using one astronomical unit as the unit of length and then the entire system is translated to a new frame using the Galilean transformation $x\mapsto x + 10^{16}$.
In the new frame, the information about the relative positions of the planets has been completely lost.
This example shows that by simply using floating point numbers, we lose Galilean invariance.
\emph{This is an unavoidable fact that is true for all integrators, not just \ias}.

In the vast majority of cases this is not a concern as long as the coordinate system is chosen with some common sense (no one would set up a coordinate system where the Solar System is centred at $10^{16}$~AU).
However, let us consider a Kozai-Lidov cycle (see below for a description of the scenario).
If a particle undergoing Kozai oscillations is offset from the origin and requires a small timestep due to its highly eccentric orbit, then the precision of the relative position during one timestep can be very low.  
The value of the $b_6$ coefficients will then be completely dominated by floating point precision, rather than the smoothness of the underlying function.
As a result, the ratio of $b_6$ to $y''$ (equation~\ref{eq:globallocal2}) is not a reliable error estimate anymore.
More importantly, the ratio does not get smaller as we further decrease the timestep.
This becomes a runaway effect as the overly pessimistic error estimate leads to a steadily decreasing timestep.
We solve this issue by not including particles in the maxima in equation~\ref{eq:globallocal2} which move very little during a timestep, i.e. particles that have a velocity smaller than $v\;dt < \alpha x$, where $\alpha$ is a small number and $x$ and $v$ are the magnitude of the particle's position and velocity during the timestep.
The results are not sensitive to the precise value of $\alpha$.
We chose to set $\alpha=10^{-8}$ to ensure that we have an effective floating point precision of roughly~8 digits in the position for the particle.
The reason for this is that $10^{-8}$ is roughly the precision needed in the $b_6$ coefficients to ensure that long term integrations follow Brouwer's law (see Section~\ref{sec:errias}).
The procedure described above is a failsafe mechanism that does not activate for well-behaved simulations such as those of the Solar System. 
It only affects simulations where the limited floating point precision does not allow for accurate force estimates.

In simulations where limited floating point precision is an issue, it is worth thinking about moving to a different, possibly accelerated frame that gives more accurate position and acceleration estimates.
The details depend strongly on the specific problem and are beyond the scope of this paper.

\subsection{Why 15th order?}
\label{sec:why}
It is natural to ask why we use a 15th-order scheme.
Although we cannot provide a precise argument against a 14th- or 16th-order scheme (this will in general be problem specific), the following considerations constitute a strong argument against schemes with orders significantly different from 15.
The physical interpretation is quite simple: we want keep the timestep at a fraction of the dynamical time, somewhere around~1\%, and we want the errors per timestep to be very small.
As it turns out, a scheme of 15th-order hits this sweet spot for calculations in double floating point precision.

Lower-order schemes will require many more timesteps per orbit to get to the required precision (recall that our scheme error after one timestep is $\sim 10^{-28}$!). 
Besides the obvious slow-down, there is another issue.
With so many more timesteps, the round-off errors will be significantly larger, since these grow with the square root of the number of timesteps.
Hence, no matter what one does, with a much lower-order scheme one simple cannot achieve the same precision as with a 15th-order scheme such as \ias.

Higher-order schemes would allow us to use even larger timesteps.
We currently use about 100 timesteps per orbit. 
So any significant increase in the accuracy of the scheme would allow us to use timesteps comparable to the dynamical time of the system. 
This become problematic for many reasons as the assumption is that the force is smooth within one timestep. 
One consequence is that a high-order scheme would have the tendancy to miss close encounters more often, simply by a lack of sampling points along the trajectory.  
We always want to resolve the orbital timescale in an N-body simulation. 

However, there is one case where higher-order schemes could be useful: simulations working in extended precision. 
To maintain a simulation accurate to within, say $10^{-34}$ (quadruple precision) instead of $10^{-16}$ (double precision), over a billion orbits, one would have to require that the error of the scheme after one timestep is less than $10^{-46}$ instead of $10^{-28}$.
Clearly, this is much easier to achieve with a higher-order scheme.

%%%%%%%%%%%%%%%%%%%%%%%%%%%%%%%%%%%%%%%%%%%%%%%%%%%%%
%%%%%%%%%%%%%%%%%%%%%%%%%%%%%%%%%%%%%%%%%%%%%%%%%%%%%
\section{Tests}
\label{sec:tests}
\begin{table*}
\caption{Integrators used in Section~\ref{sec:tests}.\label{tab:integrators}}
\begin{tabular}{l|l|l|l|l|l}
Acronym & Name/Description & Publication & Symplectic & Order& Package\\\hline \hline
\ias & Implicit integrator with Adaptive timeStepping, 15th order & this paper & No & 15 & \reb \\
WH & Wisdom-Holman mapping & \cite{WisdomHolman1991} & Yes & 2 & \reb\\
BS & Bulirsch-Stoer integrator  & \cite{BulirschStoer1966} & No & -$^\dagger$& \mer \\
RADAU & Gau{\ss}-Radau integrator  & \cite{Everhart1985} & No & 15& \mer \\
MVS & Mixed-Variable Symplectic integrator with symplectic correctors& \cite{Wisdom1996} & Yes & 2 & \mer \\
\end{tabular}

\vspace{.25cm}
\raggedright
$^\dagger$The order of the BS integrator is not given as the integrator is usually used with variable timesteps and iterated until a given convergence limit has been reached.
\end{table*}

In this section, we test our new integrator and compare it to other integrators.
For the comparison, we use integrators implemented in the \mer software package \citep{Chambers1997} and \reb \citep{ReinLiu2012}.

We chose to focus on \mer as it has proven to be very reliable and easy to use.
It also implements various different integrators and is freely available. 
For these reasons, it is heavily used by researchers in the Solar System and exoplanet community.
We also compare \ias to the Wisdom-Holman integrator (WH) of \reb.
The Wisdom-Holman integrator of \reb differs from the mixed-variable symplectic (MVS) integrator of \mer in that it does not have higher order symplectic correctors and exhibits therefore an error $E_{\rm scheme}$ about two orders of magnitude larger in a typical simulation of the Solar System.
Other integrators tested are the Bulirsch-Stoer (BS) and the \radau integrator of \mer (based on the original \citealt{Everhart1985} algorithm).
Table~\ref{tab:integrators} lists all integrators used in this section.

%%%%%%%%%%%%%%%%%%%%%%%%%%%%%%%%%%%%%%%%%%%%%%%%%%%%%
\subsection{Short term simulations of the outer Solar System}
\begin{figure}
 \centering \resizebox{3in}{!}{\includegraphics{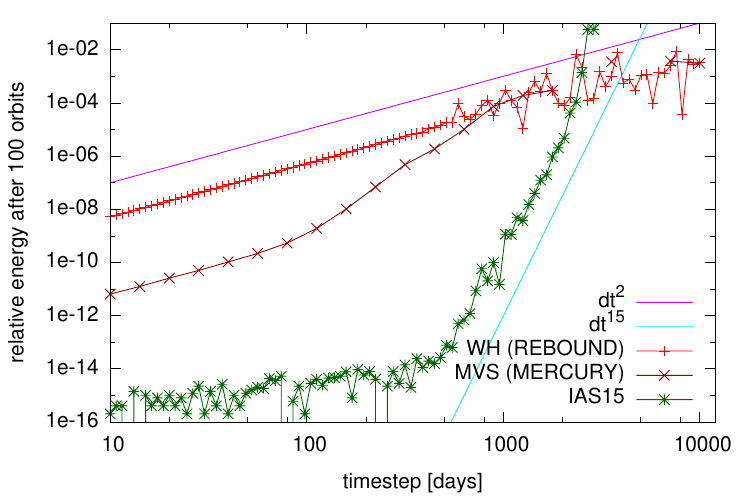}}
 \caption{
Relative energy error as a function of timestep in simulations of the outer Solar System using the \ias, WH and MVS integrator. 
Note that in this test case, \ias uses a fixed timestep.  
\label{fig:order}}
\end{figure}

\begin{figure}
 \centering \resizebox{3in}{!}{\includegraphics{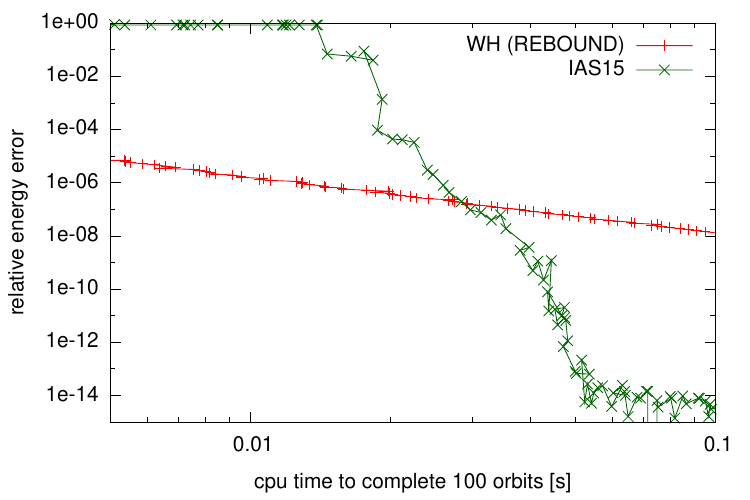}}
 \caption{
Relative energy error as a function of computation time in simulations of the outer Solar System using the \ias and WH integrator. 
Note that in this test case, \ias uses a fixed timestep.  
\label{fig:speed}}
\end{figure}

We first discuss short term simulations.
By short term, we mean maximum integration times of at most a few hundred dynamical timescales.
The purpose of these short simulations is to verify the order of \ias and compare the speed to other integrators.
 
\begin{figure}
 \centering \resizebox{3in}{!}{\includegraphics{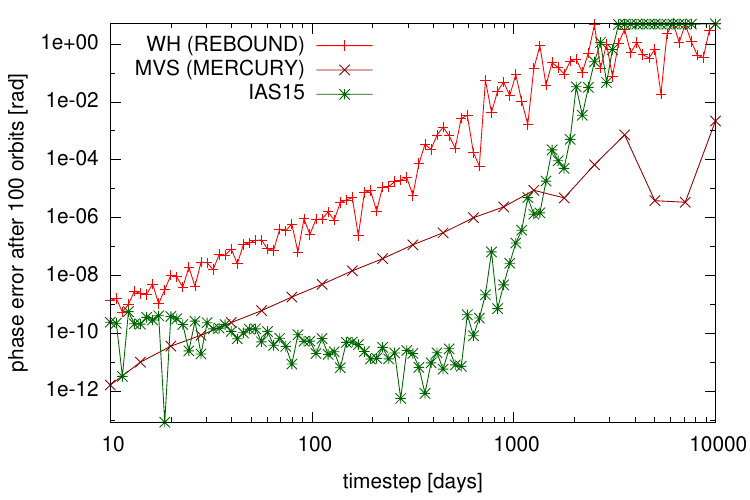}}
 \caption{
Phase error as a function of timestep in simulations of the outer Solar System using the \ias, WH and MVS integrator.
We first integrate forward 50 Jupiter periods and then the same amount backward to measure the phase offset relative to Jupiter's original position.
Note that in this test case, \ias uses a fixed timestep.  
\label{fig:phase}}
\end{figure}

In Figs.~\ref{fig:order}-\ref{fig:phase} we present results for integrations of the outer Solar System bodies (Jupiter, Saturn, Uranus, Neptune and Pluto) over 12000 years (100 Jupiter orbits).  
We measure relative energy error, phase error and the execution time while varying the timestep used in three integrators, WH, MVS and the new \ias integrator.
For this test we turn off the adaptive timestepping in \ias. 

Fig.~\ref{fig:order} shows the relative error as a function of timestep. 
From this plot, it can be verified that \ias is indeed a 15th-order integrator. 
With a timestep of $\sim 600$~days ($\sim 0.15$ Jupiter periods) \ias reaches roughly machine precision (over this integration time - see below for a complete discussion on the error over longer integration times). 
One can also verify that the WH integrator is a second order scheme.
The MVS integrator of \mer is also second order, with the error about two orders of magnitude smaller than that of WH for timesteps less than a few percent of the orbital period of Jupiter.

Note that the WH and MVS integrators have a moderate error even for very large timesteps (much larger than Jupiter's period).
It is worth pointing out that this error metric might be misleading in judging the accuracy of the scheme. 
Although WH/MVS are able to solve the Keplerian motion exactly, if the timestep is this large, any interaction terms will have errors of order unity or larger, which makes the integrators useless for resolving the non-Keplerian component of motion at these timesteps.

Fig.~\ref{fig:speed} shows the same simulations, but we plot the relative energy error at the end of the simulation versus the time it takes to complete 100~orbits.
The smaller the error and the smaller the time, the better an integrator is. 
One can see that for a desired accuracy of $10^{-6}$ or better, the \ias integrator is faster than the Wisdom-Holman integrator. 
The higher order symplectic correctors in MVS (used by \mer) improve the result as they bring down the WH error by about two orders of magnitude.\footnote{We do not show the MVS integrator in this plot as the short total integration time might overestimate the runtime of integrators in \mer.}
This shows that \ias is faster then the Wisdom-Holman integrator with symplectic correctors for a desired accuracy of $10^{-8}$ or better in this problem.

Finally, let us compare the phase error of our integrator. 
We again, integrate the outer Solar System but we now integrate it for 50 orbits into the future, then 50 orbits back in time. 
This allows us to measure the phase error very precisely.
In Fig.~\ref{fig:phase} the phase error of Jupiter is plotted as a function of the timestep for \ias, WH and MVS. 
It turns out that the \ias integrator is better at preserving the phase than WH for any timestep. 
Even when compared to the MVS integrator, \ias is better for any reasonable timestep.
The minimum phase error occurs for \ias near a timestep of 600 days ($\sim 0.15$ Jupiter periods), consistent with the minimum energy error measured above.

%%%%%%%%%%%%%%%%%%%%%%%%%%%%%%%%%%%%%%%%%%%%%%%%%%%%%
\subsection{Jupiter-grazing comets}
\begin{figure*}
 \centering 
 \resizebox{2.8in}{!}{\includegraphics{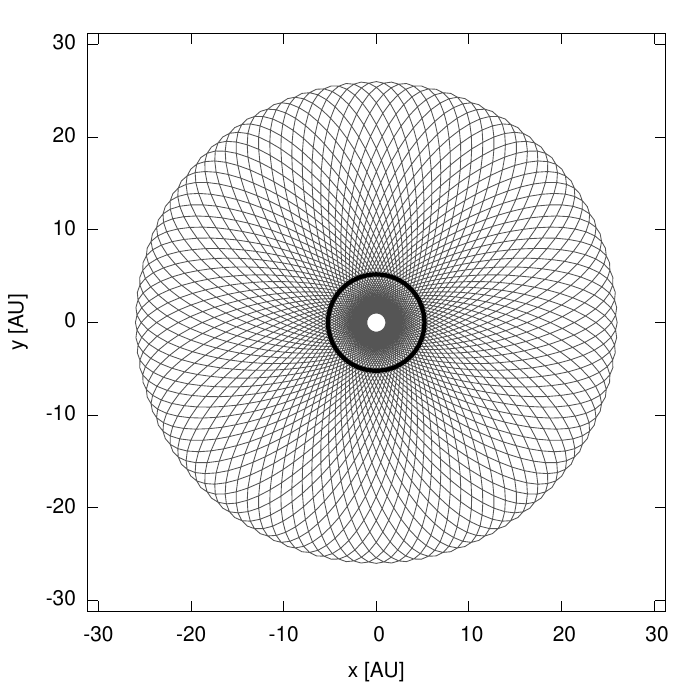}}
 \hspace{0.3in}
 \resizebox{2.8in}{!}{\includegraphics{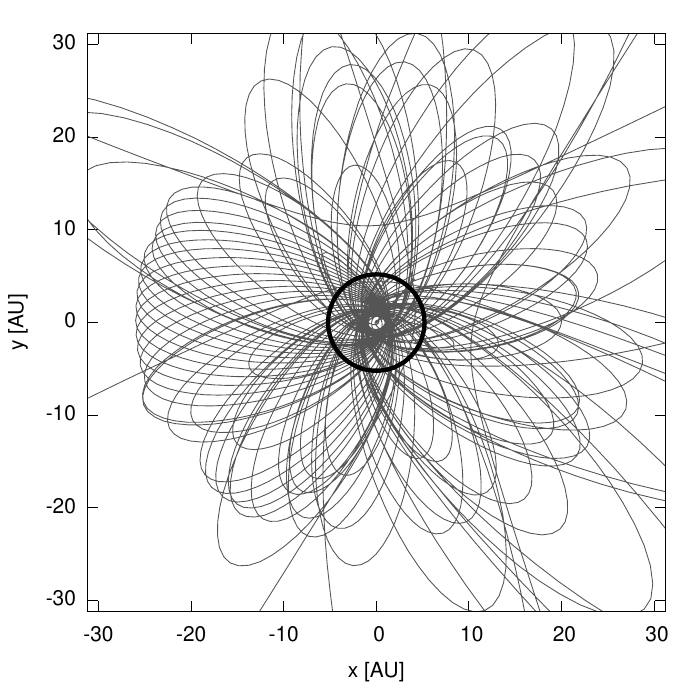}}
 \caption{
	Orbits of 100 comets crossing Jupiter's orbit. Left panel: initial orbits. Right panel: orbits after 100 Jupiter orbits. Jupiter's orbit is shown as a thick black circle.
	In this simulation the \c{maxmimum} error in the Jacobi constant is less than $10^{-14}$.
\label{fig:jacobi}}
\end{figure*}

We now move on to test the new adaptive timestepping scheme.
As a first test, we study 100 massless comets which are on paths intersecting Jupiter's orbit.
Their aphelion is at 25~AU \c{and their eccentricity is~$0.95$}.
Jupiter is placed on a circular orbit at 5.2~AU.
Thus, this is the restricted circular three body problem.

Fig.~\ref{fig:jacobi} shows the initial symmetric configuration of comet orbits as well as the final orbits after 100 Jupiter orbits and many close encounters between Jupiter and the comets. 
Throughout the simulation, the timestep is set automatically. 
We use the default value for the precision parameter of $\epsilon_b=10^{-9}$.
Every Jupiter/comet encounter is correctly detected by the timestepping scheme, resulting in a smaller timestep at each encounter. 
The closest encounter is $\sim$10 Jupiter radii. 
At the end of the simulation, the Jacobi constant\footnote{The Jacobi constant is defined as $C_J \equiv n^2 r^2 - 2(U^t + K^t)$, where $n$ is Jupiter's orbital angular velocity, $r$ is the test particle's distance from the barycentre, and $U^t$ and $K^t$ are the test particle's potential and kinetic energy as measured in the corotating system.} is preserved at a precision of $10^{-14}$ or better.
Further tests have shown that encounters within 1 to 10 Jupiter radii are still captured by the adaptive timestepping scheme.  
However, the conservation of the Jacobi constant is not as good.
\c{This is due to limited floating point precision and the choice of coordinate system (see Section~\ref{sec:errias}).}

%%%%%%%%%%%%%%%%%%%%%%%%%%%%%%%%%%%%%%%%%%%%%%%%%%%%%
\subsection{Kozai-Lidov cycles}
\label{sec:kozai}
\begin{figure*}
 \centering \resizebox{7in}{!}{\includegraphics{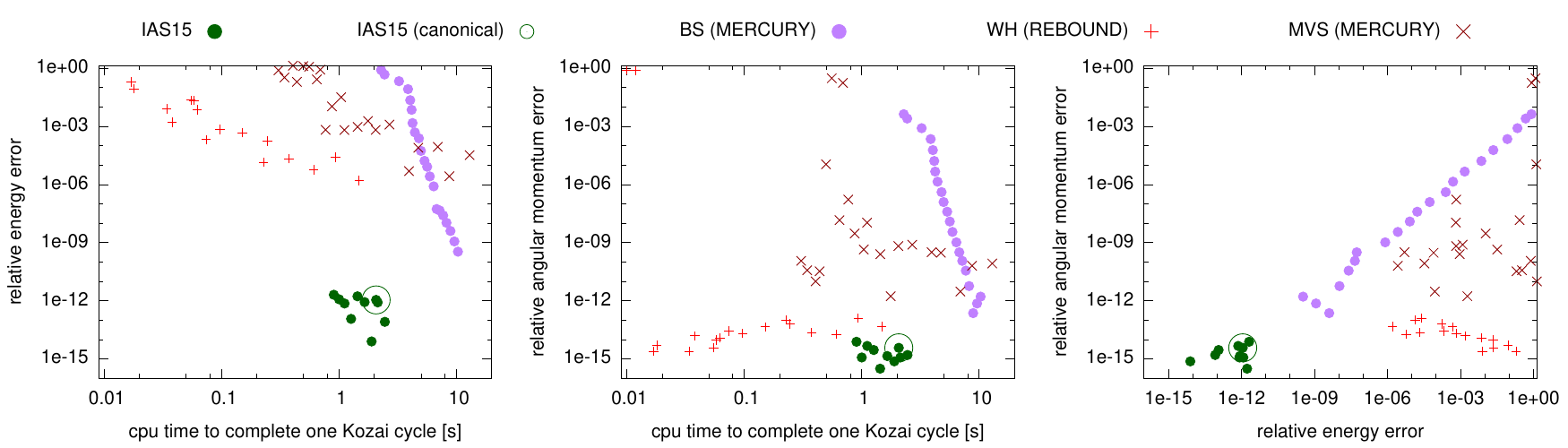}}
 \caption{
	 This plot shows results of integrations of a Kozai-Lidov cycle. 
	 The left panel shows the relative energy error as a function of CPU time.
	 The middle panel shows the relative angular momentum error as a function of CPU time.
	 The right panel shows both error metrics.
	 Results are shown for the \ias, BS, WH and MVS integrators.
\label{fig:kozai}}
\end{figure*}

In a Kozai-Lidov cycle \citep{kozai1962, lidov1962}, a particle undergoes oscillations of eccentricity and inclination on a timescale that is much longer than the orbital timescale.
These oscillations are driven by an external perturber.
Kozai-Lidov cycles can be hard to simulate due to the highly-eccentric orbits. 
We chose to use it as a test-case to verify \ias's ability to automatically choose a correct timestep.

We present results of such a setup with three particles:
a central binary with masses of $1M_\odot$ each, separated by 1~AU, and a perturber, also with mass $1M_\odot$ at a distance of 10~AU.
The inclination of the perturber's orbit with respect to the plane of inner binary orbit is $89.9^\circ$.
The highest eccentricity of the binary during one Kozai cycle is $e_{\rm max}\approx0.992$.
We integrate one full Kozai cycle, which takes about 1 second of CPU time using \ias.

In Fig.~\ref{fig:kozai} we plot the relative errors in energy and angular momentum.
The CPU time to finish the test problem is plotted on the horizontal axis for the left and middle panels.
The right panel shows the relative energy error on the horizontal axis and the relative angular momentum error on the vertical axis.
The smaller the error and the shorter the time, the better the integrator.

The results for \ias are shown in green filled circles. 
We vary the precision parameter $\epsilon_b$ over many orders of magnitude. 
The canonical value of $\epsilon_b=10^{-9}$ is shown with an open circle.
\ias preserves the energy with a precision $10^{-12}$ and the angular momentum with a precision of $10^{-15}$ in all simulations.

Note that the canonical value of $\epsilon_b$ results in the same accuracy as runs with a runtime roughly two times faster.
This suggests that we might have chosen a too conservative $\epsilon_b$, a larger value could give us equally accurate results while taking less time. 
However, we only integrated this system for one Kozai-Lidov cycle (a few thousand binary orbits).
Given that the fastest and the slowest runs are only a factor of two apart, we encourage the user to keep the default value of $\epsilon_b$, but this example shows that some further optimization could result in a small speed-up for some specific cases.

As a comparison, we also plot the results of the MVS, WH and BS integrators for a variety of timesteps and precision parameters.
None of these other integrators can handle this test case well. 
Although MVS and WH preserve the angular momentum relatively well, the energy error is large as illustrated in the right panel.
This is primarily due to the fact that the inner two bodies form an equal mass binary, a situation for which WH and MVS were not intended.
Only the BS integrator can roughly capture the dynamics of the system.
However, at equal run times, the BS integrator is several orders of magnitude less accurate than \ias as shown in the left and middle panels.

We tested many other examples of Kozai-Lidov cycles including some with extremely eccentric orbits up to $e\sim 1 - 10^{-10}$.
\ias is able to correctly integrate even these extreme cases with out-of-the-box settings.
The relative energy error in those extreme cases scales roughly as $E\sim {10^{-16}}/{(1-e_{\rm max})}$.

One important driving factor for \ias was that it should be scale independent. 
In other words, only the dynamical properties of the system should determine the outcome, not the simulation units.
We verified that \ias is indeed scale independent by varying the length and mass scales in Kozai-Lidov cycles.
It is worth pointing out that several integrators that we tested, including the \radau integrator (with either the \mer or \reb implementation) fail completely in this test problem if the system is simply rescaled.

%%%%%%%%%%%%%%%%%%%%%%%%%%%%%%%%%%%%%%%%%%%%%%%%%%%%%
\subsection{Long-term simulations}

\begin{figure*}
 \centering \resizebox{7in}{!}{\includegraphics{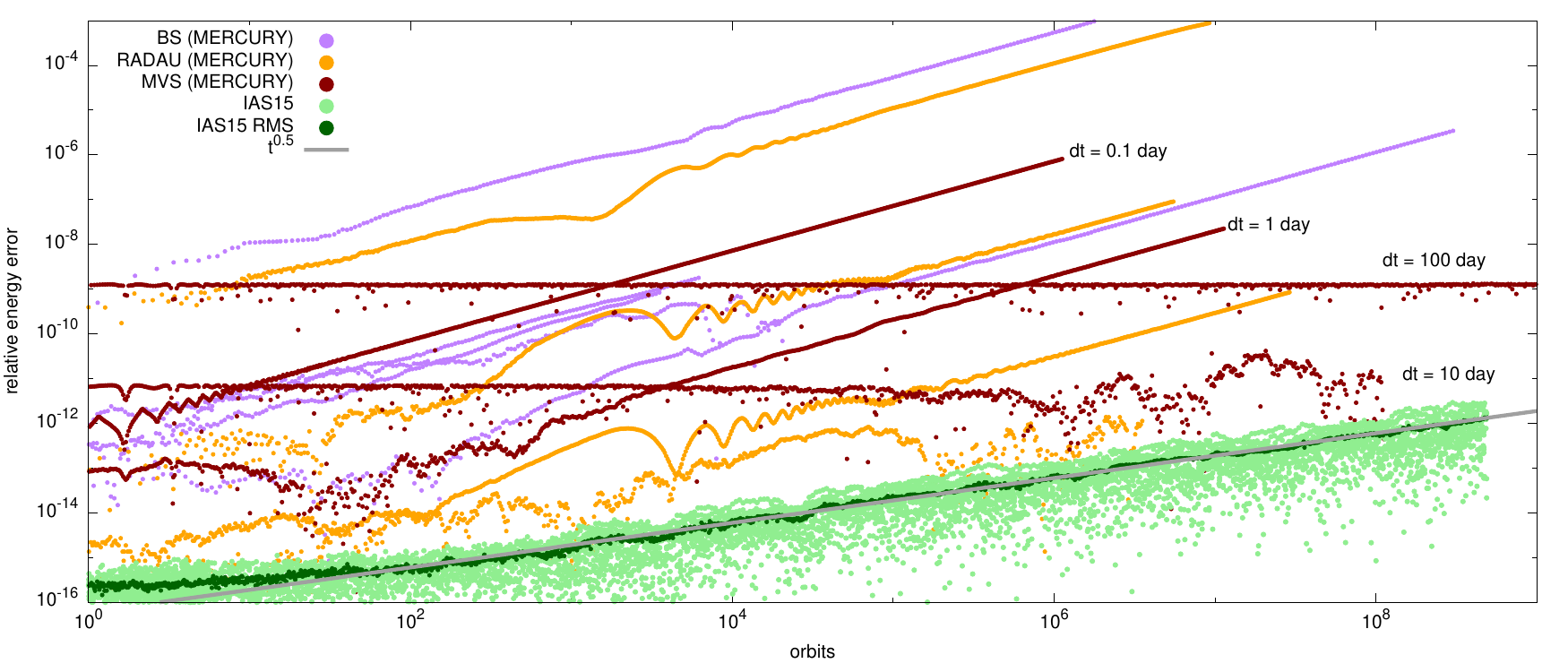}}
 \caption{
Long-term integrations of the outer Solar System.
The relative energy error is shown as a function of time (in units of Jupiter orbits) for the \ias integrator in light green for 20 different realizations of the same initial conditions.
The root mean square value of all realizations is shown in dark green.
It grows like $\sqrt{t}$ and follows Brouwer's law.
As a comparison the same quantity is plotted for the MVS, \radau and  BS integrators of \mer for a variety of timesteps and precision parameters.
\ias is more accurate than any of the other integrators shown.
MVS, despite being a symplectic integrator, shows signs of both random and linear energy growth.
The \radau integrator comes within 1 to 2 orders of magnitudes of \ias for one specific precision parameter.
However, the choice of this precision parameter is highly model dependent and fine tuned.
In other words, changing the precision parameter to larger {\it or} smaller values results in significantly larger errors.
\label{fig:longterm}}
\end{figure*}

\begin{figure*}
 \centering \resizebox{7in}{!}{\includegraphics{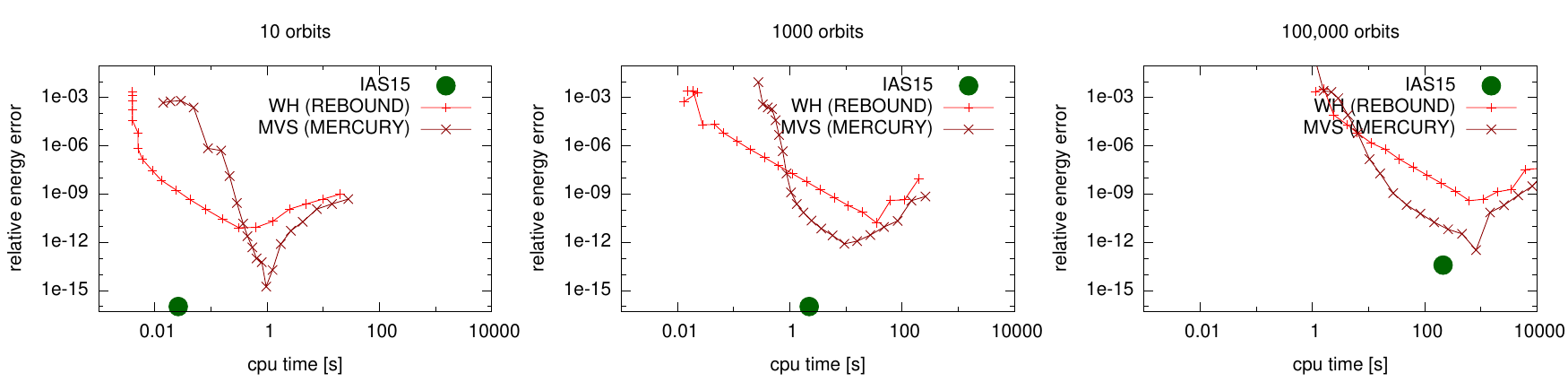}}
 \caption{
Long-term integrations.
Relative energy error as a function of run time for different integrators.
Left panel: total integration time 10 orbits.
Middle panel: total integration time 1000 orbits.
Right panel: total integration time 100,000 orbits.
\label{fig:longtermcomparison}}
\end{figure*}

Finally, let us study the behaviour of \ias in long-term simulations of planetary systems.
By long term, we mean integrations over billions of dynamical times.
As a test case, we once again chose the outer Solar System. 

Fig.~\ref{fig:longterm} shows the energy error for \ias, MVS, \radau and the Bulirsch-Stoer (BS) integrator as a function of time.
The unit of time is one Jupiter orbit, thus the plot covers nine orders of magnitude, from one to one billion dynamical timescales.
We ran MVS with different timesteps ranging from 0.1 to 100~days.
The BS and \radau integrators have adaptive timestepping. 
We ran them with various precision parameters.
Several integrations involving the BS and \radau integrators did not get past several hundred dynamical times due to very small timesteps.
To get a statistical sample, we run the \ias integrator with twenty different realizations of the same initial conditions.\footnote{We perturb the initial conditions on the level of $10^{-15}$.}
We plot both the individual tracks of \ias runs (in light green) and the root mean square value (in dark green). 

Let us first focus on \ias. 
One can easily verify that the integrator follows Brouwer's law at all times.
Initially, the error is of order of machine precision, i.e.~$E_{\rm floor}$.
After 100~orbits, random errors start to accumulate. 
We overplot a function proportional to~$\sqrt{t}$ which is an almost perfect fit to the RMS value of the \ias energy errors after 100~orbits. 
While developing the algorithm, we were able to push this growth rate down by several orders of magnitude due to 1) the optimal choice of the timestep, 2) compensated summation and 3) taking great care in implementing numerical constants.
Note that there is no linear growth in the energy error is detectable. 
This confirms that our choice of precision parameter $\epsilon_b$ was successful in making sure that $E_{\rm scheme} < E_{\rm rand}$ during the entire simulation.

Looking at the results of the BS integrator, one can see that the errors are larger than those of \ias at any time during the simulation. 
It is worth pointing out that the timesteps chosen by the BS integrator are orders of magnitudes smaller than those chosen by \ias. 
The errors are nevertheless larger because the scheme itself has a large linearly growing energy error term $E_{\rm scheme}$ associated with it. 
If this were the only error present, we would see an improvement using smaller timesteps (smaller precision parameters). 
We don't see that because of the error terms $E_{\rm bias}$ and $E_{\rm rand}$. 
The linear growth in all of the BS simulations makes it impossible to use this scheme for long-term orbit integrations.
We were not able to achieve a precision better than $10^{-8}$ after a million orbits using the BS implementation of \mer.

The error of the \radau integrator varies greatly depending on the value of the precision parameter.
Remember that this parameter is effectively a length scale, so setting it requires a priori knowledge of the system.
Even if such problem-specific fine-tuning is accomplished, the error is always at least one two to orders of magnitude larger than that of \ias.
Further reduction of the precision parameter leads, which should in principle give a more accurate result, leads in fact to a larger energy error.
Also note that the error grows linearly in time in all but one cases.
We attribute these results to an ineffective adaptive timestepping scheme, the lack of compensated summation and inaccurate representations of numerical constants.

The results for the MVS integrator look qualitatively different. 
MVS is a symplectic integrator which implies that the error term $E_{\rm scheme}$ should be bounded.
Such a behaviour can be seen in the $dt=100$~days simulation.
However, there are other error terms that are not bounded, $E_{\rm rand}$ and $E_{\rm bias}$.
The magnitude of these error terms depends on the timestep and varies from problem to problem.
This can be seen in the simulation with $dt=1$~day, where random errors dominate from the very beginning of the simulation.
An epoch of constant energy error is never reached.
After 1000~orbits, a linear energy growth can be observed, which is probably due to a bias in the implementation.
Remember that the MVS integrator needs to perform several coordinate transformations \c{(to and from Keplerian elements and barycentric and heliocentric frames)} every timestep.
This involves an iteration and calls to square root and sine functions.
At least one of these operations is biased, contributing to~$E_{\rm bias}$.
Worse yet, decreasing the timestep even more, by a factor of 10 (see $dt=0.1$~days simulation) makes the error grow faster by over a factor~100.

In summary, Fig.~\ref{fig:longterm} shows that we are unable to achieve an error as small as that achieved by the \ias simulation with either MVS, \radau or BS.
Neither using short nor long timesteps allows these other integrators to match the typical error performance of \ias on billion-orbit timescales.
\c{Another way to look at this is that \ias preserves the symplecticity of Hamiltonian systems better than the commonly-used nominally symplectic integrators to which we compared it.}
\c{Better yet, within the context of integrations on a machine with 64-bit floating-point precision, \ias~{\it is as symplectic as any integrator can possibly be.}}

Fig.~\ref{fig:longtermcomparison} illustrates this point even more, and presents another comparison of the speed of the various integrators for long-term orbit integrations.
Here, we plot the relative energy error as a function of runtime for the WH, MVS and \ias integrators.
We still study the outer Solar System.
The three panels correspond to integration times of 10, 1000 and 100,000~Jupiter orbits.
We show only one datapoint for \ias, using the default precision parameter $\epsilon_b$.
For WH and MVS we vary the timestep.

In all three panels, one can see that the error for WH and MVS initially decreases when decreasing the timestep (simulations with a smaller timestep have a longer wall time).
At some point, however, we reach an optimal timestep.
If we decrease the timestep even further, the error begins to grow again.
This is due to the terms $E_{\rm bias}$ and $E_{\rm rand}$.  
Note that neither WH nor MVS are adaptive.
It is non-trivial to find this optimal timestep for a specific problem without some experimentation.

In the first two panels, the error associated with \ias is of the order of machine precision.
In the right panel, random errors begin to show, growing as $\sqrt{t}$, consistent with Fig.~\ref{fig:longterm}.

Because \ias is always more accurate than either MVS or WH, it is hard to compare their speed directly (i.e. MVS and WH can never get as accurate as \ias, no matter how long we wait).
What we can compare \ias to is the most accurate simulation of MVS and WH. 
Even in that case, \ias is always faster than either MVS or WH.

%%%%%%%%%%%%%%%%%%%%%%%%%%%%%%%%%%%%%%%%%%%%%%%%%%%%%
%%%%%%%%%%%%%%%%%%%%%%%%%%%%%%%%%%%%%%%%%%%%%%%%%%%%%
\section{Conclusions}
\label{sec:conclusions}
In this paper we have presented \ias, a highly-accurate, 15th-order integrator for N-body simulations.
\ias is based on the \gr integration scheme of \citet{Everhart1985} and features a number of significant improvements, including compensated summation to minimize loss of precision in floating-point operations, and a new, physically-motivated, dynamic timestepping algorithm that is robust against rescaling the problem.

In Section~\ref{sec:errors}, we discussed different types of errors that can afflict a celestial mechanics integration.
These errors include $E_{\rm floor}$, owing to the finite numerical precision with which floating-point numbers are stored in a digital computer; $E_{\rm rand}$, the random component of longterm accumulation of floating-point errors; $E_{\rm bias}$, the non-random part associated with longterm accumulation of floating-point errors; and $E_{\rm scheme}$, the error associated with a particular integration method.
We examine the error behaviour of \ias and find that $E_{\rm bias}$ and $E_{\rm scheme}$ remain subdominant even over an integration covering a billion dynamical timescales.

Although there is no unambiguous definition of \emph{goodness} when comparing N-body integrators, using a wide range of metrics, \ias is better than all other integrators that we tested in a wide range of astrophysically interesting problems.
These metrics include errors in energy, phase, angular momentum, Jacobi constant, and the runtime required to achieve a desired error level.
We have taken great care in implementing \ias.
Most importantly, we have made use of compensated summation, and we have optimized numerical constants to ensure proper rounding of floating-point operations.
Because of this, we achieve an energy error that follows Brouwer's law in simulations of the outer Solar System over at least $10^9$ dynamical timescales~(12~billion~years). 
In this sense, we call \ias \emph{optimal}. 
Its error is limited only by the limits of floating-point precision.
It is not possible to further improve the accuracy of the integration without using some sort of extended floating-point precision. 

Other groups have published integrators that they have claimed to be optimal.
Although we do not have access to the implementation used by \cite{Grazier2005}, comparing their fig.~2 to our Fig.~\ref{fig:longterm} shows that \ias conserves the energy better at a given number of orbits by $\sim$1-2 orders of magnitude.
%Note: it is a small "f" in fig when the figure is in another paper.
Part of this difference might be because \ias uses a significantly larger timestep.
\cite{Hairer2008} present an implicit Runge-Kutta scheme that achieves Brouwer's law, but tests have been conducted over much shorter timescales.

Our comparison in this paper includes symplectic integrators which are often seen as desirable because their energy errors (in problems that can be described by a Hamiltonian) are generally considered to be bounded.
\ias is not symplectic.
Despite this, \ias has better long-term error performance (by a variety of metrics) than symplectic integrators such as leapfrog, the mixed-variable symplectic integrator of \citet{WisdomHolman1991}, or an MVS integrator with higher-order correctors \citep{Wisdom1996}.
\c{Thus, we conclude that \ias preserves the symplecticity of Hamiltonian systems better than nominally symplectic integrators do.\footnote{One could go as far as calling \ias itself a symplectic integrator.}}
Furthermore, symplectic integrators are by construction not well suited for integrations of non-Hamiltonian systems, where velocity-dependent or other non-conservative forces, such as Poynting-Robertson (PR) drag, are present.
The error performance of \ias, however, is not affected in these situations.

Mixed-variable symplectic integrators work typically in the heliocentric frame. 
Whereas this makes sense for integrations of planetary systems, it is a non-ideal choice for other problems, such as those involving central binaries (see Section~\ref{sec:kozai}). 
\ias does not require a specific frame and can therefore handle a much wider range of astrophysical problems.

Maybe most importantly and in addition to all the benefits already mentioned above, \ias is an excellent out-of-the-box choice for practically all dynamical problems in an astrophysical context.
The default settings will give extremely accurate results and an almost optimal runtime.
Fine tuning the integrator for a specific astrophysical problems can result in a small speedup.
However, this fine tuning can render the integrator less robust in other situations.
In all the testing we have done, we have found that \ias, with out-of-the-box settings and no fine-tuning, is either the best choice by any metric, or at least not slower by more than a factor of $\sim$2-3 than any fine-tuned integrator.
Furthermore, although some other integrators can achieve comparable error performance in some specific test cases, these integrators can fail spectacularly in other cases, while \ias performs extremely well in all the cases we looked at.

To make \ias an easy-to-use, first-choice integrator for a wide class of celestial mechanics problems, we provide a set of astrophysically interesting test problems that can easily be modified.
These including problems involving Kozai-Lidov cycles, objects with nonzero quadrupole moment, radiation forces such as~PR~drag~(with optional shadowing), and non-conservative migration forces.
\ias is freely available within the \reb package at \url{https://github.com/hannorein/rebound}.
\c{We provide both an implementation in C99 and a python wrapper.}

\vspace{0.5in}

\section*{Acknowledgments}
We are greatly indebted to Scott Tremaine for many useful discussions and suggestions.
We thank the referee, John Chambers, for a careful review that improved the manuscript.
HR and DSS both gratefully acknowledge support from NSF grant AST-0807444.
DSS acknowledges support from the Keck Fellowship and from the Association of Members of the Institute for Advanced Study.

\bibliography{full}

\begin{thebibliography}{36}
\expandafter\ifx\csname natexlab\endcsname\relax\def\natexlab#1{#1}\fi

\bibitem[{{Brouwer}(1937)}]{Brouwer1937}
{Brouwer}, D. 1937, \aj, 46, 149

\bibitem[{Bulirsch \& Stoer(1966)}]{BulirschStoer1966}
Bulirsch, R. \& Stoer, J. 1966, Numerische Mathematik, 8, 1

\bibitem[{{Burns} {et~al.}(1979){Burns}, {Lamy}, \& {Soter}}]{Burns1979}
{Burns}, J.~A., {Lamy}, P.~L., \& {Soter}, S. 1979, \icarus, 40, 1

\bibitem[{{Chambers}(1999)}]{chambers1999}
{Chambers}, J.~E. 1999, \mnras, 304, 793

\bibitem[{{Chambers} \& {Migliorini}(1997)}]{Chambers1997}
{Chambers}, J.~E. \& {Migliorini}, F. 1997, in Bulletin of the American
  Astronomical Society, Vol.~29, AAS/Division for Planetary Sciences Meeting
  Abstracts \#29, 1024

\bibitem[{{Everhart}(1985)}]{Everhart1985}
{Everhart}, E. 1985, in Dynamics of Comets: Their Origin and Evolution,
  Proceedings of IAU Colloq 83, ed. A.~{Carusi} \& G.~B. {Valsecchi}, Vol. 115,
  185

\bibitem[{Feng(1985)}]{Feng1985}
Feng, K. 1985, Proceedings of the 1984 Beijing Symposium on Differential
  Geometry and Differential Equations, 42

\bibitem[{Fousse {et~al.}(2007)Fousse, Hanrot, Lef\`evre, P\'elissier, \&
  Zimmermann}]{Fousse2007}
Fousse, L., Hanrot, G., Lef\`evre, V., P\'elissier, P., \& Zimmermann, P. 2007,
  {ACM} Transactions on Mathematical Software, 33, 13:1

\bibitem[{{Gladman} {et~al.}(1991){Gladman}, {Duncan}, \&
  {Candy}}]{Gladman1991}
{Gladman}, B., {Duncan}, M., \& {Candy}, J. 1991, Celestial Mechanics and
  Dynamical Astronomy, 52, 221

\bibitem[{Grazier {et~al.}(2005)Grazier, Newman, Hyman, Sharp, \&
  Goldstein}]{Grazier2005}
Grazier, K.~R., Newman, W.~I., Hyman, J.~M., Sharp, P.~W., \& Goldstein, D.~J.
  2005, in Proc. of 12th Computational Techniques and Applications Conference
  CTAC-2004, ed. R.~May \& A.~J. Roberts, Vol.~46, C786--C804

\bibitem[{{Hairer} {et~al.}(2006){Hairer}, {Lubich}, \& {Wanner}}]{Hairer2006}
{Hairer}, E., {Lubich}, C., \& {Wanner}, G. 2006, {Geometric Numerical
  Integration} (Springer)

\bibitem[{Hairer {et~al.}(2008)Hairer, McLachlan, \& Razakarivony}]{Hairer2008}
Hairer, E., McLachlan, R.~I., \& Razakarivony, A. 2008, BIT Numerical
  Mathematics, 48, 231

\bibitem[{{H{\'e}non} \& {Petit}(1998)}]{HenonPetit1998}
{H{\'e}non}, M. \& {Petit}, J.-M. 1998, Journal of Computational Physics, 146,
  420

\bibitem[{Henrici(1962)}]{Henrici1962}
Henrici, P. 1962, Discrete variable methods in ordinary differential equations
  (Wiley)

\bibitem[{Higham(2002)}]{Higham2002}
Higham, N. 2002, Accuracy and Stability of Numerical Algorithms: Second Edition
  (Society for Industrial and Applied Mathematics)

\bibitem[{{Hildebrand}(1974)}]{Hildebrand1974}
{Hildebrand}, F.~B. 1974, {Introduction to numerical analysis} ({McGraw-Hill,
  New York})

\bibitem[{Kahan(1965)}]{Kahan1965}
Kahan, W. 1965, Commun. ACM, 8, 40

\bibitem[{{Kozai}(1962)}]{kozai1962}
{Kozai}, Y. 1962, \aj, 67, 591

\bibitem[{{Kronrod}(1965)}]{Kronrod1965}
{Kronrod}, A.~S. 1965, {Nodes and weights of quadrature formulas. Sixteen-place
  tables} (Consultants Bureau New York, Authorized translation from Russian)

\bibitem[{{Lidov}(1962)}]{lidov1962}
{Lidov}, M.~L. 1962, \planss, 9, 719

\bibitem[{{Malhotra}(1994)}]{Malhotra1994}
{Malhotra}, R. 1994, Celestial Mechanics and Dynamical Astronomy, 60, 373

\bibitem[{{Mikkola}(1997)}]{Mikkola1997}
{Mikkola}, S. 1997, Celestial Mechanics and Dynamical Astronomy, 68, 249

\bibitem[{{Newcomb}(1899)}]{Newcomb1899}
{Newcomb}, S. 1899, Astronomische Nachrichten, 148, 321

\bibitem[{{Poynting}(1903)}]{Poynting1903}
{Poynting}, J.~H. 1903, \mnras, 64, A1

\bibitem[{{Quinlan} \& {Tremaine}(1990)}]{quinlan+tremaine1990}
{Quinlan}, G.~D. \& {Tremaine}, S. 1990, \aj, 100, 1694

\bibitem[{{Quinn} \& {Tremaine}(1990)}]{QuinnTremaine1990}
{Quinn}, T. \& {Tremaine}, S. 1990, \aj, 99, 1016

\bibitem[{{Rein} \& {Liu}(2012)}]{ReinLiu2012}
{Rein}, H. \& {Liu}, S.-F. 2012, \aap, 537, A128

\bibitem[{{Rein} \& {Tremaine}(2011)}]{ReinTremaine2011}
{Rein}, H. \& {Tremaine}, S. 2011, \mnras, 845+

\bibitem[{{Robertson}(1937)}]{Robertson1937}
{Robertson}, H.~P. 1937, \mnras, 97, 423

\bibitem[{Ruth(1983)}]{Ruth1983}
Ruth, R.~D. 1983, IEEE Transactions on Nuclear Science, 30, 2669

\bibitem[{{Schlesinger}(1917)}]{Schlesinger1917}
{Schlesinger}, F. 1917, \aj, 30, 183

\bibitem[{{Tricarico}(2012)}]{ORSA}
{Tricarico}, P. 2012, {ORSA: Orbit Reconstruction, Simulation and Analysis},
  astrophysics Source Code Library

\bibitem[{Vogelaere(1956)}]{Vogelaere1956}
Vogelaere, R. 1956, Methods of integration which preserve the contact
  transformation property of the Hamiltonian equations (University of Notre
  Dame)

\bibitem[{Weyl(1939)}]{Weyl1939}
Weyl, H. 1939, Princeton mathematical series, Vol.~1, The Classical Groups:
  Their Invariants and Representations (Princeton University Press)

\bibitem[{{Wisdom} \& {Holman}(1991)}]{WisdomHolman1991}
{Wisdom}, J. \& {Holman}, M. 1991, \aj, 102, 1528

\bibitem[{{Wisdom} {et~al.}(1996){Wisdom}, {Holman}, \& {Touma}}]{Wisdom1996}
{Wisdom}, J., {Holman}, M., \& {Touma}, J. 1996, Fields Institute
  Communications, Vol.~10, p.~217, 10, 217

\end{thebibliography}

\begin{appendix}
\section{Simple Derivation of Poynting-Robertson drag}
\label{sec:PR}
We provide an example problem within \reb that uses our new \ias integrator to integrate a non-Hamiltonian system including Poynting-Robertson drag.
\c{The relevant files can be found in the directory \texttt{examples/prdrag/}.}
In this appendix, we provide what we consider an intuitive and simple derivation of this non-conservative force.

Consider a dust particle orbiting a star on a circular orbit.
In the following we will assume that the dust particle absorbs every photon from star and reemits it in a random direction.
Let the dust particle's velocity be $\vec{\mathbf{v}}$ and its position be $r \hat{\mathbf{r}}$, where $r$ is the distance from the star and $\hat {\mathbf{r}}$ the unit vector pointing from the star to the particle.
The radiation forces felt by the dust particle can be described as a simple drag force between the dust particle and photons from the star.

Following \citet{Burns1979}, we say that the radiation pressure force that would be experienced by a stationary dust particle at this position is a factor $\beta$ smaller than the gravitational force from the star, and define
\begin{eqnarray}
  F_r \equiv \frac{\beta G M_*}{r^2} \, .
  \label{eq:Fr}
\end{eqnarray}
Here, $\beta \sim 3L_*/(8\pi c \rho G M_* d)$, where $L_*$ is the star's luminosity, $c$ is the speed of light, $\rho$ is the density of the dust grain, $G$ is the gravitational constant, $M_*$ is the star's mass, and $d$ is the diameter of the dust grain.  Note that, since $F_r$ scales as $r^2$, radiation pressure acts in a way that is effectively equivalent to reducing the mass of the star by a factor of $\beta$.

We now move on to non-stationary particles.
Let us first consider a particle moving radially,
\begin{eqnarray}
\vec{\mathbf{v}}_{\rm radial} = \dot{r}  \hat{\mathbf{r}}.
\end{eqnarray}
Taking this relative movement between star and dust particle into account, we can calculate the radial component of the radiation force,
\begin{eqnarray}
\vec{\mathbf{F}}_{\rm radial} = F_r \brp{1 - 2\frac{\dot{r}}{c}} \hat{\mathbf{r}} \, .
\end{eqnarray}
There are two contributions to the term involving $\dot{r}/c$, hence the coefficient of 2.
Firstly, the energy per photon is Doppler boosted by a factor of $1-\dot{r}/c$.
Secondly, the dust particle moves at radial speed $\dot{r}$ relative to the star and therefore encounters photons a factor of $1-\dot{r}/c$ more often.
Bringing the two contributions together, the force is increased by $\brp{1 - \dot{r}/c}^2 \approx 1 - 2\dot{r}/c$ for $|\dot r| \ll c$.

The radial force is not the only force felt by an orbiting dust particle.
There is also an azimuthal component, which depends on the azimuthal velocity,
\begin{eqnarray}
\vec{\mathbf{v}}_{\rm azimuthal} = \vec{\mathbf{v}} - \vec{\mathbf{v}}_{\rm radial} \, .
\end{eqnarray}
For $v \equiv |\mathbf{v}|\ll c$, the azimuthal component is just the radial force scaled by the ratio of the azimuthal velocity to the speed of light,
\begin{eqnarray}
\label{eq:Ftan0} \vec{\mathbf{F}}_{\rm azimuthal} &=& F_r \frac{-\vec{\mathbf{v}}_{\rm azimuthal}}{c} +O\left[(v/c)^2\right]\, ,
\end{eqnarray}
which may be rewritten as
\begin{eqnarray}
\nonumber\vec{\mathbf{F}}_{\rm azimuthal} & = & F_r \frac{ -(\vec{\mathbf{v}} - \vec{\mathbf{v}}_{\rm radial}) }{c} \\
\label{eq:Ftan} & = & F_r \frac{ \dot{r}\hat{\mathbf{r}} - \vec{\mathbf{v}}  }{c} \, .
\end{eqnarray}
The total force is then the sum of the radial and azimuthal components
\begin{eqnarray}
\nonumber \vec{\mathbf{F}}_{\rm photon} & = & \vec{\mathbf{F}}_{\rm radial} + \vec{\mathbf{F}}_{\rm azimuthal} 
  = F_r \left\{ \left(1 - 2\frac{\dot{r}}{c}\right) \hat{\mathbf{r}} + \frac{ \dot{r}\hat{\mathbf{r}} - \vec{\mathbf{v}}  }{c} \right\} \\
 & = & F_r \left\{ \left(1 - \frac{\dot{r}}{c}\right) \hat{\mathbf{r}} - \frac{ \vec{\mathbf{v}}  }{c} \right\}
\end{eqnarray}
Even though this force is due to just one physical effect (photons hitting a dust particle), the different components are often referred to as different processes.
The term ``radiation pressure'' is used for the radial component and the term ``Poynting-Robertson drag'' is used for the azimuthal component.

\end{appendix}

\end{document}